\documentclass[%
 aip,
 amsmath,amssymb,
 reprint,%
]{revtex4-1}

\usepackage{graphicx}
\usepackage{dcolumn}
\usepackage{bm}

\usepackage[utf8]{inputenc}
\usepackage[T1]{fontenc}
\usepackage{mathptmx}
\usepackage{etoolbox}

\makeatletter
\def\@email#1#2{%
 \endgroup
 \patchcmd{\titleblock@produce}
  {\frontmatter@RRAPformat}
  {\frontmatter@RRAPformat{\produce@RRAP{*#1\href{mailto:#2}{#2}}}\frontmatter@RRAPformat}
  {}{}
}%
\makeatother
\begin{document}

\preprint{}

\title[Nanoscale Water Behavior and Its Impact on Adsorption]{Nanoscale Water Behavior and Its Impact on Adsorption: \\ A case study with CNTs and Diclofenac}
\author{Patrick R. B. Côrtes}
\affiliation{Departamento de Física, Instituto de Física e Matemática, Universidade Federal de Pelotas, Caixa Postal 354, 96.001-970, Pelotas, Brasil}
\author{Nicolás A. Loubet}
\author{Cintia A. Menéndez}
\author{Gustavo A. Appignanesi}
\affiliation{INQUISUR, Departamento de Química, Universidad Nacional del Sur (UNS)-CONICET, Avenida Alem 1253, 8000 Bahía Blanca, Argentina}
\author{Mateus H. K\"ohler}
\email{mateus.kohler@ufsm.br}
\affiliation{Departamento de F{\'\i}sica, Universidade Federal de Santa Maria, 97105-900, Santa Maria, Brasil}
\author{José Rafael Bordin}
\email{jrbordin@ufpel.edu.br}
\affiliation{Departamento de Física, Instituto de Física e Matemática, Universidade Federal de Pelotas, Caixa Postal 354, Pelotas, Brasil}

\date{\today}

\begin{abstract}
 Water is a fundamental component of life, playing a critical role in regulating metabolic processes and facilitating the dissolution and transport of essential molecules. However, the presence of emerging contaminants, such as pharmaceuticals, poses significant challenges to water quality and safety. Nanomaterials-based technologies arise as a promising tool to remove those contaminants from water. Nevertheless, interfacial water plays a major role in the adsorption of chemical compounds in the nanomaterials - as it plays in biological processes such as protein folding, enzyme activity, and drug delivery. To understand this role, in this study we employ Molecular Dynamics (MD) simulations to explore the adsorption dynamics of potassium diclofenac (K-DCF) on single-walled (SWCNT) and double-walled (DWCNT) carbon nanotubes, considering both dry and wet conditions. Our findings reveal that the structuring of water molecules around CNTs creates hydration layers that significantly influence the accessibility of active sites and the interaction strength between contaminants and adsorbents. Our analysis indicates higher energy barriers for adsorption in DWCNTs compared to SWCNTs, which is attributed to stronger water-surface interactions. This research highlights the importance of understanding nanoscale water behavior for optimizing the design and functionality of nanomaterials for water purification. These findings can guide the development of more efficient and selective nanomaterials, enhancing contaminant removal and ensuring safer water resources, while also contributing to a deeper understanding of fundamental biological interactions.
\end{abstract}

\maketitle

\section{Introduction}

Water is indispensable for life, serving as a fundamental component in regulating metabolic processes. It acts as a solvent, facilitating the dissolution and transport of nutrients, gases, and waste products throughout living systems~\cite{laage2017,bordin2023}. In biological systems, the relevance of water extends to the adsorption process, which is vital for the proper function of cellular membranes and enzymes. Water molecules form hydration shells around biomolecules, influencing their structure, stability, and interactions~\cite{kohler2017protein,Khodadadi2017,Shikata2024}. This hydration layer is crucial for the adsorption of substrates on enzyme active sites, affecting the efficiency of metabolic reactions and nutrient absorption~\cite{Mondal2022}. Additionally, hydration water affects the adsorption of drugs by biomolecules, influencing their bioavailability and therapeutic effectiveness by contributing to the free-energy of the interaction between the drug molecule and their targets~\cite{biman05,fenter2014protein}. Therefore, understanding the behavior of water in adsorption processes is essential for advancing our knowledge of life~\cite{weng2019,Pal2021}.

Also, ensuring safe drinking water for everyone is a significant challenge. To address this and other emerging social issues, in 2015, the United Nations signed a pact with several countries to establish 17 Sustainable Development Goals (SDGs) to be achieved by 2030~\cite{Agenda2023}. Two of these goals are directly linked to water quality: SDG 6, "Clean Water and Sanitation," and SDG 14, "Life Below Water." However, recent reports indicate that we are not on track to meet these goals~\cite{SDG2022}.

An even greater challenge is dealing with contaminants of emerging concern~\cite{kumar2022review,geissen2015emerging,bell2011emerging,deblonde2011emerging}. Some pharmaceuticals, when dispersed in water, can be categorized as pollutants that not only render water unfit for consumption but also directly impact aquatic life~\cite{klimaszyk2018water,jorgensen2000drugs}, as they are designed to elicit physiological responses in living organisms~\cite{reichert2019emerging,Wang2021}.

Non-steroidal anti-inflammatories are of particular interest because they are widely available at a relatively low cost, which, combined with erroneous disposal of medicines and packaging, leads to the pollution of canals and river basins~\cite{Svanfelt2010,Lin2023,Ellepola2022}. One of the most consumed non-steroidal anti-inflammatories is diclofenac (DCF). Moreover, it has been found that traditional water treatment processes are ineffective in removing this compound from water. As a consequence, it is often detected in the effluents from plants and in receiving water bodies. Studies indicate that DCF is extremely toxic to many animal species~\cite{Ajima2014,dalligan2021,HerreroVillar2021,Trombini2020,wiacka2020,Joachim2021,Wolf2021}. For instance, diclofenac has been found to cause renal failure in vultures, leading to a significant decline in their populations in regions where the drug is used in veterinary medicine~\cite{green04,Oaks2004}. In aquatic environments, DCF exposure has been linked to physiological and reproductive harm in fish, including kidney damage and reduced fertility~\cite{Schwaiger2004,Triebskorn2004}. Other aquatic organisms, such as invertebrates and amphibians, are also vulnerable to the toxic effects of DCF, which can disrupt their endocrine systems and negatively impact their populations~\cite{Mohd2017,Han2010}.

Nanomaterials-based technologies for water purification are promising solutions to address the growing concern of water contamination by emerging pollutants~\cite{leao2023a,bordin2024review,deazevedo2023comprehensive}. Water molecules form hydration layers around nanomaterials and pollutants, significantly influencing the adsorption dynamics~\cite{schewe2021shell,suzuki1997}. These hydration layers affect the accessibility of active sites on nanomaterials, the diffusion of pollutants, and the overall interaction strength between pollutants and adsorbents. Therefore, a detailed understanding of water behavior at the nanoscale is crucial for optimizing the design and functionality of nanomaterials for water purification. By studying these interactions, researchers can develop more efficient and selective nanomaterials, ultimately improving the removal of contaminants and ensuring safer water resources for various applications.

Carbon-based nanomaterials, such as carbon nanotubes (CNTs) and graphene, stand out as promising candidates for developing new technologies for water decontamination, particularly in the removal of emerging contaminants~\cite{Yin2019}. These nanomaterials exhibit exceptional properties, including high surface area, chemical stability, and tunable surface functionalities, which make them highly effective adsorbents for a wide range of pollutants~\cite{bordin2024review,Thakur2024,Hsu2023}. The unique structure of carbon-based nanomaterials allows for strong interactions with various contaminants, including pharmaceuticals, pesticides, and heavy metals, thereby enhancing their removal efficiency from water systems~\cite{Manimegalai2023}. Then, studying the adsorption of diclofenac in carbon nanotubes is particularly relevant as it helps to understand the mechanisms and factors influencing the adsorption efficiency and capacity of CNTs, guiding the optimization and design of more effective CNT-based nanomaterials~\cite{Mishra2023}.

Computer simulations emerge in this scenario as an alternative to access information at an atomic level and for the rational design of nanomaterials with specific properties for water treatment~\cite{Li2015}. Specifically, Molecular Dynamics (MD) simulations have been widely used to study the adsorption of many emerging contaminants on carbon-based materials~\cite{cortes2024unraveling,ibrahim2024adsorption,yang2021efficient,araujo2024,leao2023b}. More recently, the influence of the solvent in processes taking place in aqueous solutions has been investigated~\cite{moscato2024unraveling,chow2023nuclear,franco2023theoretical,bux2020solvation}. This is even more remarkable at the solid-liquid interface, especially considering the unique behavior of confined water, whose structure is quite different from that of bulk water~\cite{kohler2019water,ho2014molecular,kohler2016size}.

In this direction, we use MD simulations to study how the molecular structure of water near carbon nanotubes (CNTs) impacts the adsorption of emerging contaminants. The potassium diclofenac (K-DCF) molecule was chosen to represent the contaminant. We compared the adsorption efficiency of single-walled (SWCNT) and double-walled (DWCNT) carbon nanotubes in two scenarios: dry, when the CNT interior is not filled with water, and wet, when the CNT interior is filled with water. Our findings indicate that the combination of water structuration and the number of CNT layers plays an important role, explaining the differences between both systems. Our paper is organized as follows: in Section 2, we present the simulation details and methods. In Section 3, the results are discussed, and in Section 4, we highlight the main conclusions and remarks.

\section{The Model and Simulation Details}

\begin{figure*}[ht]
    \centering
    \includegraphics[scale=0.3]{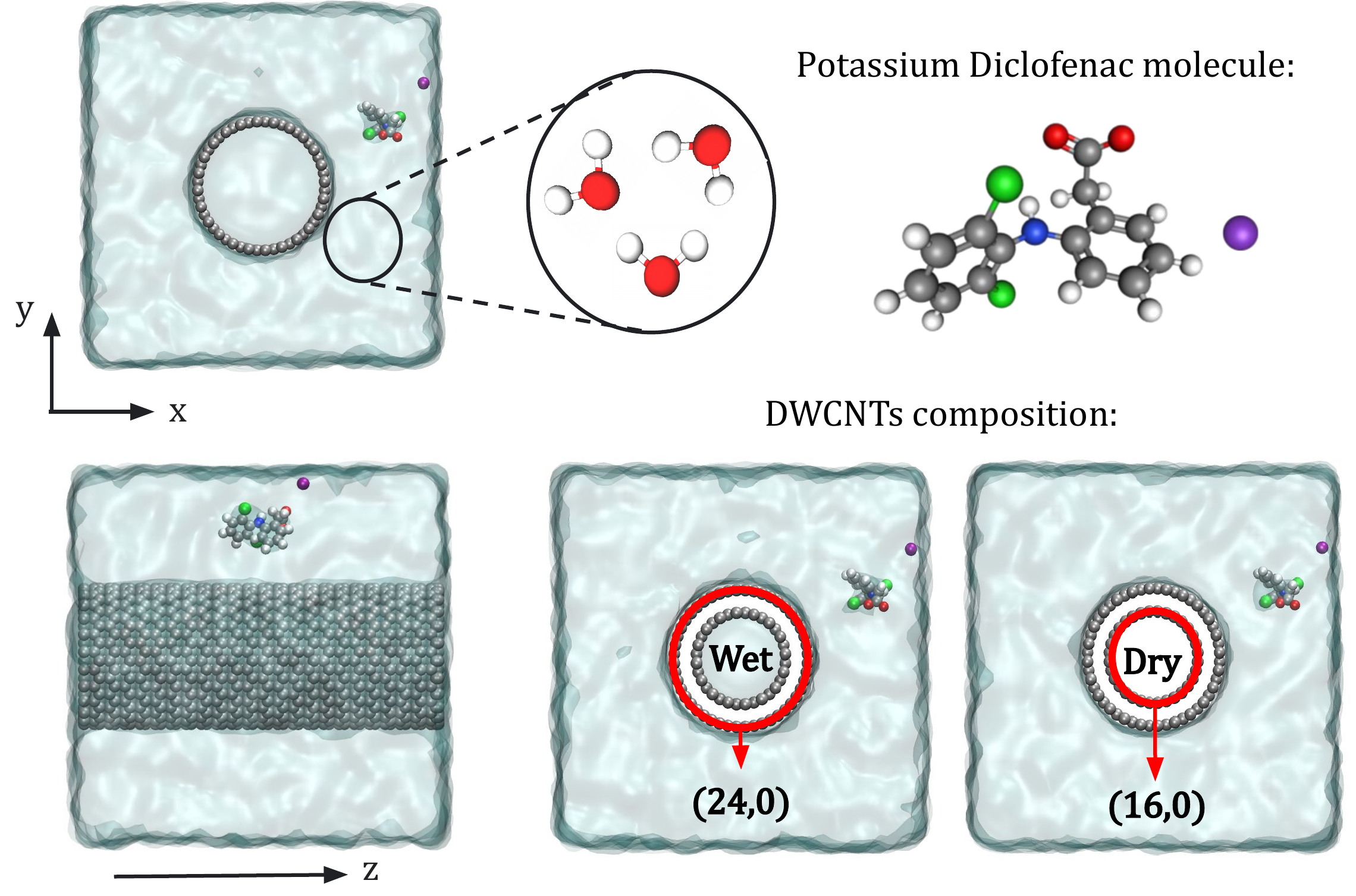}
    \caption{Depictions of simulated systems, diclofenac molecule, and CNTs orientation. The color scheme for atoms follows: red is Oxygen, white is Hydrogen, gray is Carbon, green is Chloride, blue is Nitrogen, and purple is the Potassium ion.}
    \label{fig:system}
\end{figure*}

The simulated systems, as illustrated in Figure \ref{fig:system}, consist of a single- or double-walled nanotube fixed at the center of the box, a potassium diclofenac molecule, and water. The simulation setup was built using the Moltemplate~\cite{jewett2021moltemplate} software. We use a 5 nm long zigzag (24,0) nanotube for the SWCNT system and a (24,0) enclosing another (16,0) nanotube for the DWCNT system, both also with 5 nm in length. The chiralities were chosen to reproduce the inter-wall nanotube gap found in the literature~\cite{lu1997elastic}. The CNT Lennard-Jones (LJ) parameters were obtained from the work by Saito et al. \cite{saito2001anomalous}

The potassium diclofenac structure was downloaded from the Protein Data Bank \cite{berman2000protein} (PDB) and kept rigid during the simulation. The LJ parameters and charges were based on the work of Levina et al.~\cite{levina2018hydration} for the diclofenac and the sodium parameters as parametrized by Fuentes-Azcatl and Barbosa~\cite{fuentes2018potassium}. The water was modeled as SPC/E water~\cite{berendsen1987missing}, and Lorentz-Berthelot mixing rules were employed. 4100 water molecules were accommodated in a $5 \times 5 \times 5$ nm simulation box to achieve the bulk water density 1 g/cm$^3$, and the temperature was kept constant at 300 K using the Nosé-Hoover thermostat~\cite{nose1984molecular,hoover1985canonical}. 

All simulations were performed using the LAMMPS package ~\cite{thompson2022lammps}. They all started with an energy minimization step, followed by 1 ns of thermalization and then 10 nanoseconds of data accumulation.
Periodic boundary conditions were applied. Molecules were maintained rigid with the RATTLE algorithm~\cite{andersen1983rattle}. Three sets of independent simulations, with different initial positions and velocity distributions, were performed for each configuration (SWCNT and DWCNT with or without water inside).

To analyze the trajectories of the DCF molecules, we chose to map out the Nitrogen coordinates through the simulation time. With the radial distance distributions between the DCF and the outer walls of the CNT we could establish adsorption criteria. The water radial density profile from the nanotube was also computed to assess a possible route involving the water-mediated DCF adsorption in CNTs. Additionally, the Potential of Mean Force (PMF)~\cite{abal2021molecular,kohler2018cation,bordin2012ion} was calculated for all systems.

To quantify the hydrophobicity level of the nanotubes, we calculate the probability $P(N)$ of finding N water molecules inside a sphere of radius 3.3 {\AA} tangent to the CNT. Larger values of $P(N=0)$ indicate higher hydrophobicity, which is inversely proportional to the work necessary to completely remove water molecules from the observation volume. In this scenario, the quantity $-k_B T \ln{P(N=0)}$ can be understood as the energy spent to create a cavity tangent to the nanotube~\cite{loubet2024water} and can be related to the energy barrier confronted by the DCF molecule to occupy that same cavity.

Recently, the $V_{4S}$ index was proposed~\cite{loubet2024water,loubet2023structural}. It accounts for energetic information at the atomic level in an MD simulation being calculated by establishing four interaction sites in the water molecule and estimating the potential energy due to each of them. Once we are using the SPC/E water model, the first two sites are located at the positions of hydrogen atoms. They are 1 {\AA} away from the oxygen, forming an H-O-H angle of $109.47^\circ$, the tetrahedral angle. The last two sites are determined to complete a perfect tetrahedron, i.e., 1 {\AA} away from the oxygen on the opposite side, producing the same angle as H-O-H but in an orthogonal plane to the first plane formed by the atoms of the water molecule. Four tetrahedral interaction sites are established this way. Having determined the 4 points, we calculate the potential energy between the water molecule and each heavy atom surrounding it and then attribute the contributions to the nearest site. After adding up all contributions at each site, we have four energy interaction values that we organize from the highest in absolute value, that is, the most negative one ($V_{1S}$) to the lowest contribution ($V_{4S}$). More details can be found in the very recent works~\cite{loubet2024water,loubet2023structural,montes2020structural}, while a programming code to implement this indicator can be found in:
https://github.com/nicolas-loubet/V4S.

\section{Results and Discussions}

\begin{figure*}[ht]
    \centering
    \includegraphics[scale=0.45]{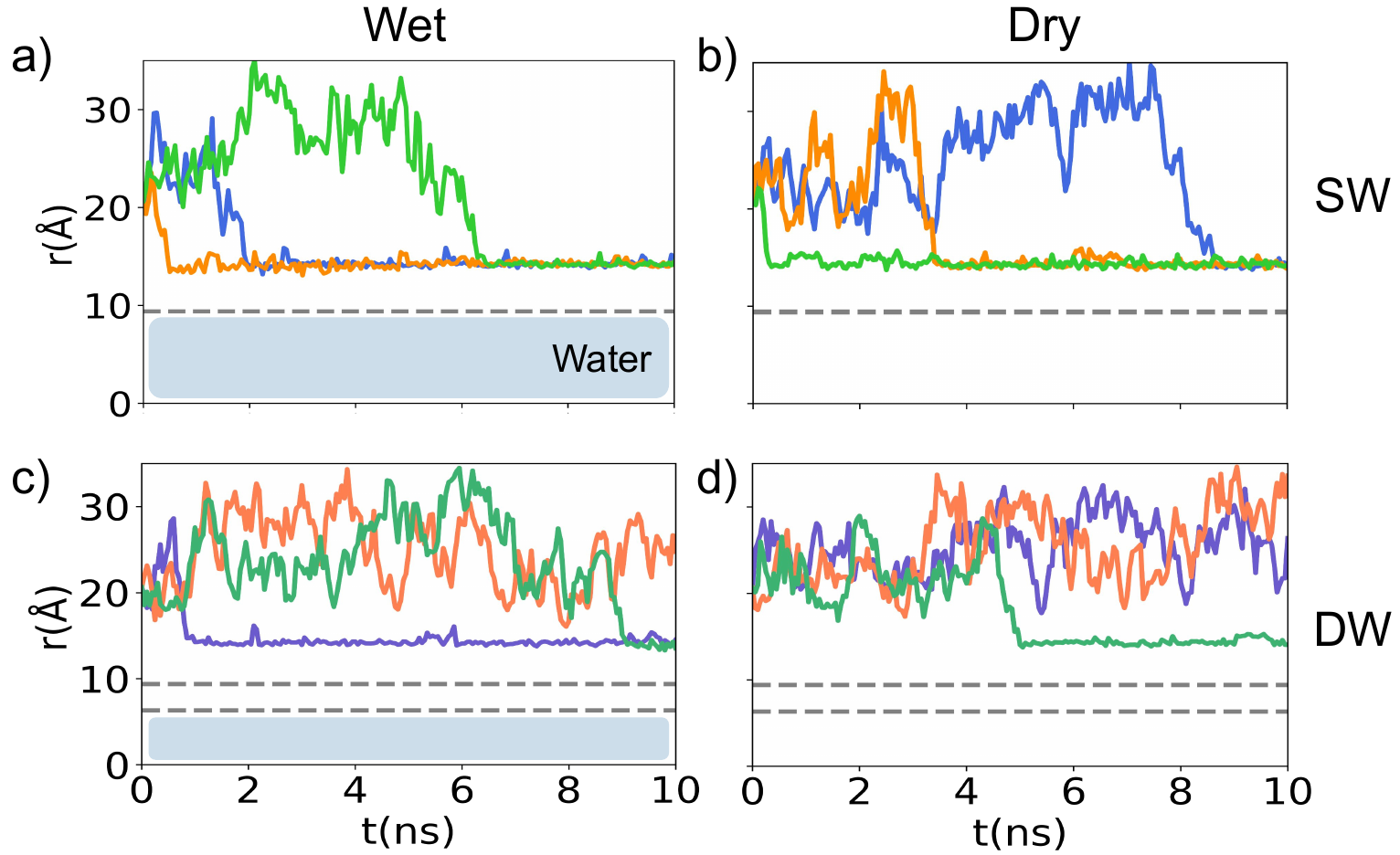}
    \caption{Radial position of Diclofenac molecule over time in the four cases simulated. Each color line represents one round of simulation.}
    \label{fig:radial}
\end{figure*}

To estimate the adsorption of the compound in the CNT we evaluate the radial distance $ r = \sqrt{x^2+y^2} $ of the k-DCF molecule relative to the center of the CNT, as shown in Figure \ref{fig:radial}. The dashed line indicates the position of the nanotube wall, and each line represents the radial trajectories of the molecule from three distinct simulations. The trajectories in Figure \ref{fig:radial}(a) and (b) indicate that for single-walled CNTs, the k-DCF molecule performs a random walk, with both shorter (less than 1 ns) and longer walks (approximately 8 ns), until it encounters the CNT wall. It was observed that once diclofenac reaches this minimum distance from the CNT, it remains at this radial distance until the end of the simulation, indicating adsorption. However, for DWCTN simulations, adsorption appears more challenging.

For wet DWCNTs, three distinct cases were observed: in the first case (blue curve in Figure \ref{fig:radial}(c)), the molecule is adsorbed at the beginning of the simulation (approximately 1 ns). In the second simulation (green curve), adsorption occurs at the end of the simulation (approximately 1 ns), and in the third simulation (orange curve), the k-DCF molecule does not reach the contact distance during the simulation time. On the other hand, adsorption in dry DWCNTs was observed in one of the simulations, as shown in Figure \ref{fig:radial}(d).

\begin{figure}[ht]
    \centering
        \includegraphics[width=0.4\textwidth]{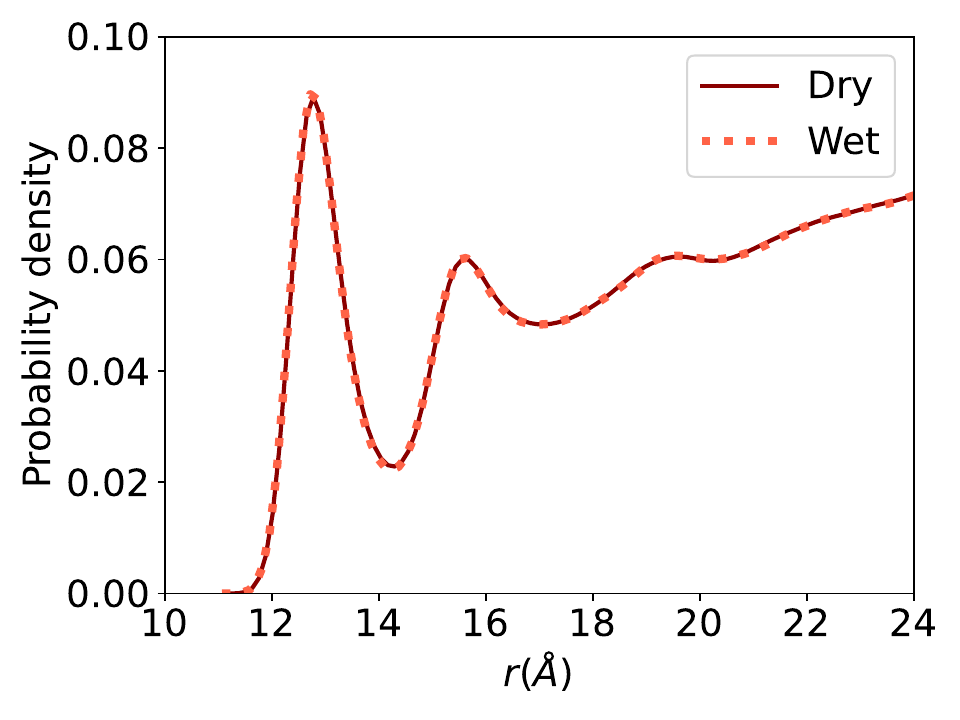}
        \includegraphics[width=0.4\textwidth]{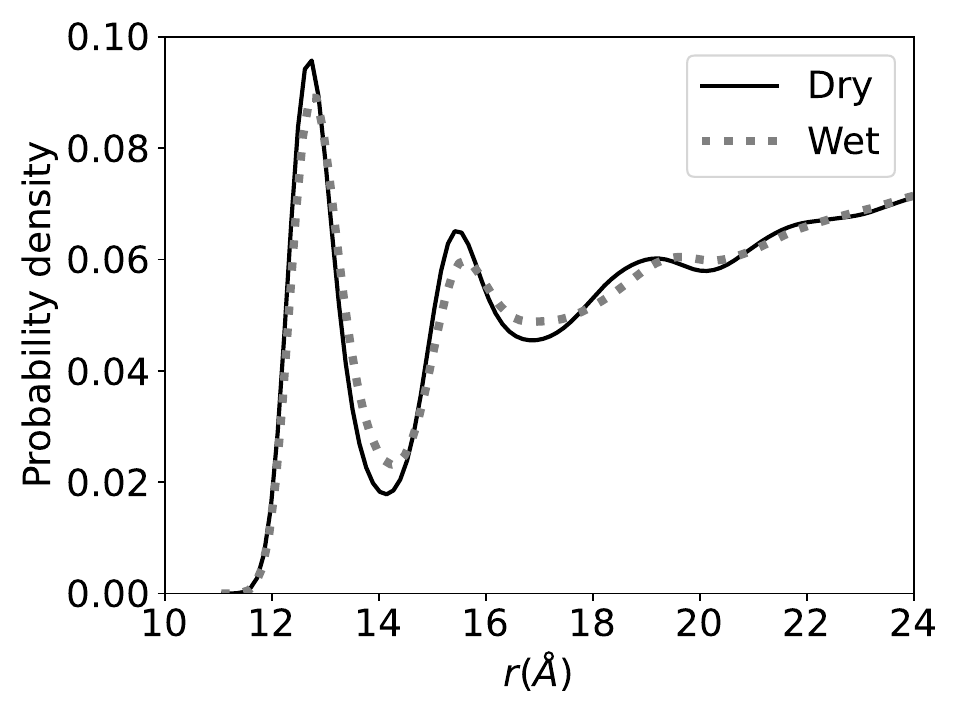}
    \caption{Probability density to find an Oxygen atom of a water molecule at a certain distance from CNTs for SWCNT (a) and (b) DWCNT.}
    \label{fig:probabilidades}
\end{figure}

\begin{figure}[ht]
     \centering
        \includegraphics[width=0.4\textwidth]{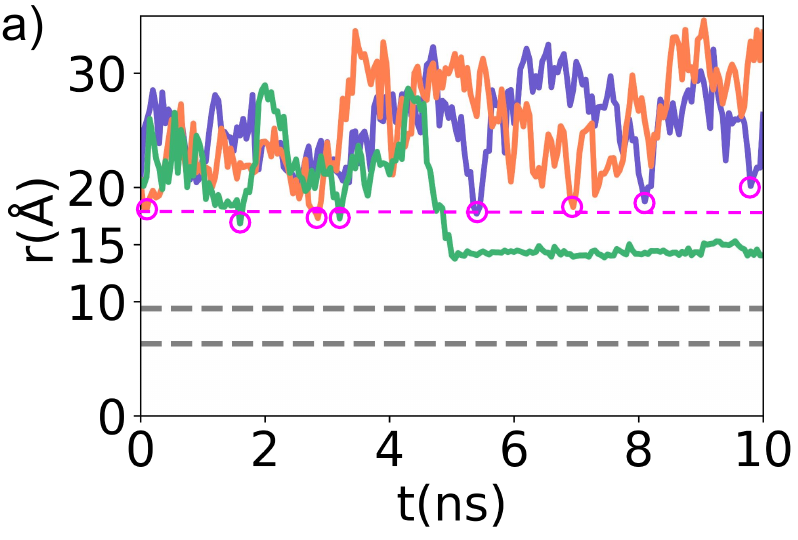}
        \includegraphics[width=0.4\textwidth]{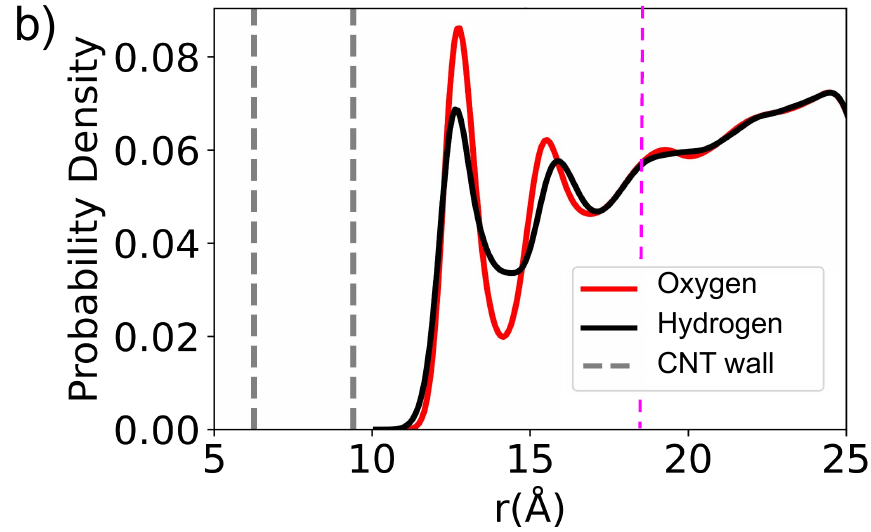}
   
     \caption{(a) Qualitative interpolation of return point of k-DCF molecule in DWCNT dry system and b) Probability Density to find an atom of a water molecule at a given distance. The pink dashed lines are only guides pointing to the pink circles interpolation in a) and the end of the structuration in (b).}
    \label{fig:barrier}
\end{figure}

It is unclear whether longer simulations would lead to all drug molecules being adsorbed by the CNTs and how this would happen regardless of the number of walls or whether they are filled with water. However, it is clear that adsorption is more challenging in DWCNTs. Interestingly, when molecules are not adsorbed, they appear to encounter a barrier and bounce back. To clarify this, let's take a closer look at the dry DWCNT case. 

As seen in Figure \ref{fig:barrier}(a), even in cases where the molecule was eventually adsorbed, it initially reaches a minimal distance of $r \approx 18$ {\AA} and bounces back before overcoming this barrier and adsorbing. The magenta circles indicate points where the blue trajectory curve reaches the location of this barrier, marked by the dashed magenta line in Figure \ref{fig:barrier}(a).

To understand the origin of this barrier, we can analyze the water structure around the nanotube in this case. Figure \ref{fig:barrier}(b) shows the probability density of water molecules in the radial direction, highlighting the two layers structure of water around the CNT. By comparing Figures \ref{fig:barrier}(a) and (b), we can relate the barrier near 18 Å to the water's two layer structure. In other words, the k-DCF molecule must overcome this barrier created by the water structure around the CNT. Since adsorption was observed in all cases with SWCNTs but only in half of the DWCNT simulations, it suggests that the water structural barrier is higher near DWCNTs than SWCNTs.

However, by comparing the probability densities of finding water molecules near SWCNTs and DWCNTs, as shown in Figure \ref{fig:probabilidades}, no clear differences in water distribution around the distinct nanotubes are observed. In all cases, two layers of water molecules are observed, indicating that this parameter alone is insufficient to distinguish the water structuring around different nanotubes. To better understand these differences, it is essential to analyze the energy and tetrahedrality of the water molecules. By examining the energy interactions, we can gain insights into the stability and binding strength of the water molecules with the nanotube surfaces. Additionally, evaluating the tetrahedrality of water molecules can help us understand their local structural organization. Tetrahedrality measures the degree to which water molecules form tetrahedral arrangements, a hallmark of their hydrogen-bonding network. By combining these analyses, we can better understand the water molecule structure and behavior around SWCNTs and DWCNTs, providing a clearer picture of how different nanotube structures influence water molecule arrangement and interactions.

\begin{figure*}[t]
    \centering
        \includegraphics[width=0.45\textwidth]{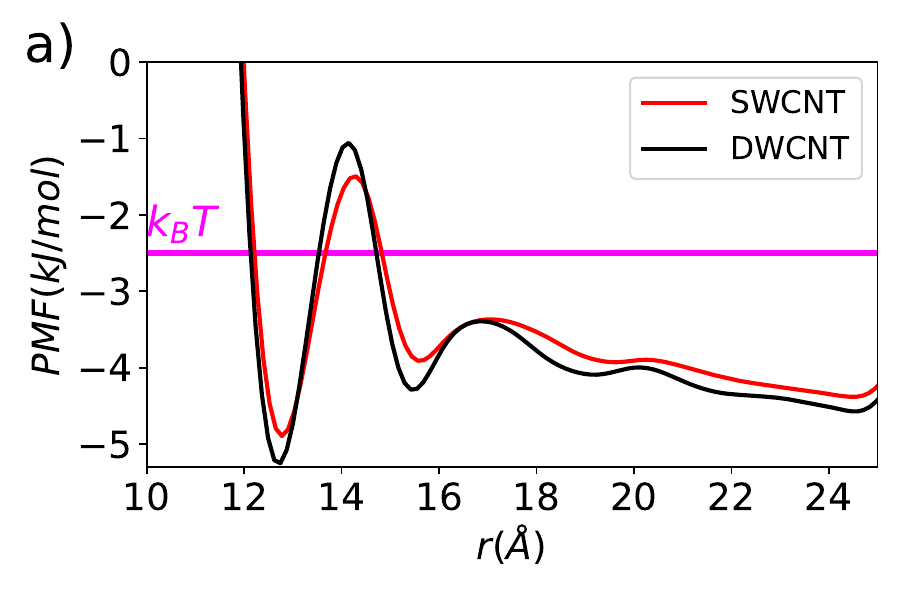}
        \includegraphics[width=0.45\textwidth]{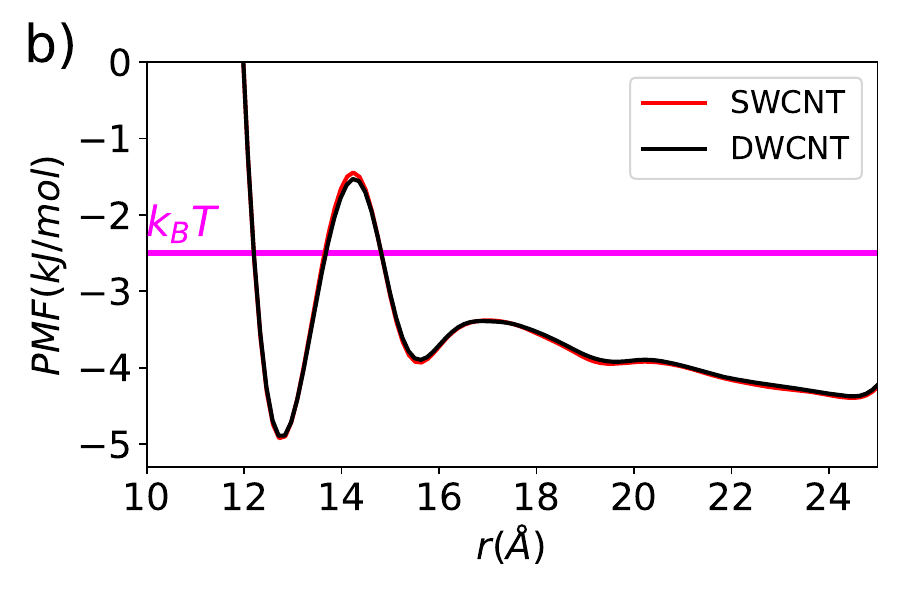}
        \includegraphics[width=0.45\textwidth]{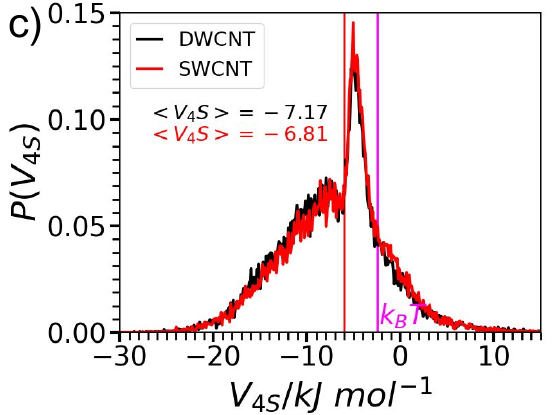}
        \includegraphics[width=0.45\textwidth]{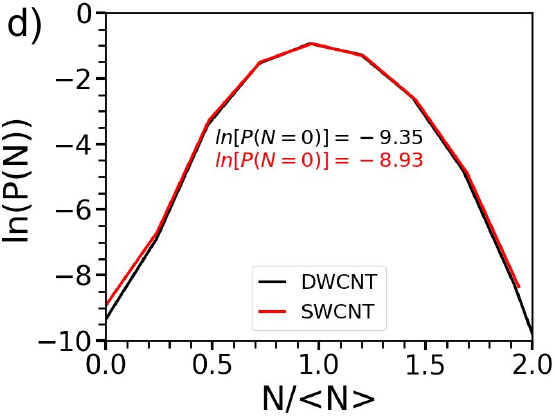}
    \caption{Potential of Mean Force around for a) dry and b) hydrated CNT nanotubes, c) Probability of finding N water molecules inside the observation sphere, comparations for SWCNT and DWCNT systems and, d) Distribution of $V_{4S}$.}
    \label{fig:energia}
\end{figure*}

The energy barrier felt by the k-DCF molecule, created by the water structure, is shown in Figure \ref{fig:energia}(a) and (b) by the radial PMF around the nanotube. It provides insight into the most favorable radial positions for a solute molecule, indicated by the valleys. It's interesting to notice that for the dry CNTs there is a significant difference for the SWCNT and DWCNT PMFs, see Figure \ref{fig:energia}(a), while for completely hydrated CNTs the number of walls does not play a significant role, as Figure \ref{fig:energia}(b) indicates. It indicates that, for the dry case, the carbons from the inner wall affect the water structure on the outside, while when the CNT is fully hydrated, this effect vanishes. This result indicates that, within the analyzed systems, a molecule will face more difficulty approaching the double-walled, dry nanotube.

This behavior is explained by the relationship of the $V_{4S}$ indices presented in Figure \ref{fig:energia}(c). Here, we explore only the dry case to see if any difference can be observed. The $V_{4S}$ index is determined as the smallest, in magnitude, of the potential energies of the four interaction points related to the tetrahedral arrangement of the water molecule. This means that, since water molecules have a lower average $V_{4S}$ around DWCNTs than SWCNTs (from $-7.17$ to $-6.81$), the interactions of water molecules are stronger in the former case. This result is consistent with the findings shown in Figure \ref{fig:energia}(c), which quantifies the hydrophobicity of the nanotube by calculating the probability of finding N water molecules inside a small observation sphere tangent to the nanotube wall. The differences in $\ln(P(N=0))$, with a slightly lower value for the DWCNT, indicate greater work required to create a vacuum cavity at the water-nanotube interface. Indeed, we note that this 5\% difference in ln(P(N=0) translates itself into a rougly 50\% increase in the probability of cavity creation, P(N=0), at the SWCNT as compared to the DWCNT, pointing to the fact that it is easier to remove hydration water from the SWCNT. Consistently, the difference in the mean value of V4s for the two nanotubes also amounts to around 5\%. Such a difference arises from the fact that a water molecule close to the DWCNT interacts with a greater number of C atoms as compared to the SWCNT (even when the interaction with the inner wall of the DWCNT is much atenuated by its distance). In turn, since $V_{4S}$ senses the water-wall interaction (which should be overcome in order to remove the water molecule), it also implies that the SWCNT presents around 50\% higher dehydration propensity.

In summary, the differences in the behavior observed in Figure \ref{fig:radial} are explained by subtle differences in the structuring of water around the nanotubes, where the number of walls and the existence or not of water inside the CNTs can affect the adsorption process. The return point in the DWCNT systems is explained both by the energy barriers for transitioning between the first solvation layers, denoted by the PMF peaks in Figure \ref{fig:energia}a, and by the greater difficulty in in removing water molecules around the double-walled nanotubes compared to the SWCNT, allowing the diclofenac molecule to be adsorbed.

\section{Conclusion}

In this study, we utilized MD simulations to investigate the impact of water molecular structure near CNTs on the adsorption of K-DCF, a prominent example of emerging contaminants. Our results reveal that the structuring of water molecules and the number of CNT walls significantly influence the adsorption dynamics.
We observed that water molecules form hydration layers around the CNTs, which affect the accessibility of active sites and the strength of interaction between pollutants and adsorbents. In single-walled nanotubes, K-DCF molecules exhibited a random walk until adsorption occurred, while in the double-walled case the adsorption process was more complex and often hindered by an energy barrier created by the structured water layers.

The energy barrier for adsorption in DWCNTs, highlighted by the Potential of Mean Force (PMF) analysis, was higher compared to SWCNTs. This difference is attributed to the stronger interaction of water molecules with DWCNTs, as indicated by the lower $V_{4S}$ indices and higher work required to create a vacuum cavity at the water-nanotube interface. These subtle differences in water structuring and energetics explain the observed adsorption behaviors and highlight the importance of considering nanoscale water behavior in the design of nanomaterials for water purification.

Our findings emphasize that both the structural characteristics of CNTs and the presence of confined water play crucial roles in the adsorption efficiency of contaminants. This knowledge can guide the development of more effective and selective nanomaterials, ultimately enhancing the removal of pollutants and ensuring safer water resources. Future research should continue to explore the interplay between water structuring and nanomaterial properties to optimize water treatment technologies. Additionally, the approach employed in this work has broader implications for understanding the role of water in biological processes. For instance, hydration layers are critical in protein folding, where water influences the structural stability and function of proteins. Similarly, water molecules around active sites affect substrate binding and catalytic efficiency in enzyme activity. By studying water behavior in these contexts, we could gain insights into fundamental biological processes such as signal transduction, where water mediates interactions between biomolecules, and drug delivery, where hydration layers influence the bioavailability and efficacy of therapeutic agents. Thus, the insights from this study would not only be helpful in advancing water purification technologies but might also enhance our understanding of the central role of water in various biological processes.

\section*{Acknowledgement}

Without public funding this research would be impossible. J. R. B. thanks the National Council for Scientific and Technological Development (CNPq), under grant numbers 405479/2023-9, 441728/2023-5, and 304958/2022-0, as well as from the Rio Grande do Sul Research Foundation (FAPERGS), grant number 21/2551-0002024-5.  M. H. K. thanks CNPq Universal grant number 306709/2021-0 and FAPERGS PqG Grant number 21/2551-0002023-7. P. R. B. C. acknowledge support from the Coordination for the Improvement of Higher Education Personnel Brasil (CAPES), Financing Code 001. C.A.M., N.A.L. and G.A.A. gratefully acknowledge funds from UNS and CONICET.

\nocite{*}
\bibliography{references}

\end{document}